\documentclass[11pt, a4paper]{article}

\usepackage{color}
\usepackage{graphicx}
\usepackage{amssymb}
\usepackage{amsmath}
\usepackage[english]{babel}
\usepackage[T1]{fontenc}
\usepackage[noadjust]{cite}%per le citazioni multiple del tipo [1-3] in luogo di [1, 2, 3]
\usepackage{hyperref} %carica i collegamenti ipertestuali all'interno del testo e con file esterni: utile per arrivare direttamente al pdf di qualche articolo in arXiv

\usepackage[utf8]{inputenc}
\usepackage{authblk}

\newcommand\eeq{\end{equation}}
\newcommand\beq{\begin{equation}}
\newcommand\eea{\end{eqnarray}}
\newcommand\bea{\begin{eqnarray}}

\begin{document}

\linespread{1.1}

\title{ \color{red} \bf Ruling out Critical Higgs Inflation?}

\author[1,2,3]{ {\Large Isabella Masina} \thanks{masina@fe.infn.it}}

\affil[1]{\small Dip. di Fisica e Scienze della Terra, Ferrara University and INFN, Ferrara, Italy }
\affil[2]{${\rm CP}^{3}$ - Origins \& DIAS, Southern Denmark University, Odense, Denmark}
\affil[3]{Theoretical Physics Department, CERN, Geneva, Switzerland}
\date{}

\maketitle

\begin{abstract}
We consider critical Higgs inflation, namely Higgs inflation with a rising inflection point at smaller field values than those of the plateau induced by the non-minimal coupling to gravity. It has been proposed that such configuration is compatible with the present CMB observational constraints on inflation, and also with primordial black hole production accounting for the totality or a fraction of the observed dark matter. We study the model taking into account the NNLO corrections to the Higgs effective potential: such corrections are extremely important to reduce the theoretical error associated to the calculation. 
We find that, in the $3\,\sigma$ window for the relevant low energy parameters, which are the strong coupling and the Higgs mass (the top mass follows by requiring an inflection point), the potential at the inflection point is so large (and so is the Hubble constant during inflation) that the present bound on the tensor-to-scalar ratio is violated. 
The model is viable only allowing the strong coupling to take its upper $3-4\, \sigma$ value. 
In our opinion, this tension shows that the model of critical Higgs inflation is likely to be not viable: neither inflation nor black holes as dark matter can be originated in this version of the model. 
%Generalizing the model by the introduction of right-handed neutrinos, the tension does not disappear.
\end{abstract}

\linespread{1.2}

\vskip 1.cm
%%%%%%%%%%%%%%%%%%%%%%%%
\section{Introduction}

Critical Higgs inflation \cite{Bezcritical, Hamada:2014iga, Hamada:2014wna} is a particular case of Higgs inflation \cite{BezHiggs, Beztwoloop}, in which the Higgs potential displays a rising inflection point at field values just below the plateau induced by the non-minimal coupling to gravity, $\xi$. It was introduced with the peculiarity of accounting for quite large, potentially observable, tensor-to-scalar ratio of cosmological perturbations, $r$, with respect to the standard scenario of Higgs inflation with a large non-minimal coupling $\xi$, where $r$ is predicted to be approximately $0.003$, together with a scalar tilt of curvature perturbation $n_s \approx 0.97$ \cite{BezHiggs}.  
Previous analysis of critical Higgs inflation \cite{Bezcritical, Hamada:2014iga, Hamada:2014wna} exploited some approximated form for the effective potential, without discussing in detail the theoretical error associated to such approximation:
their aim was primarily to show that $r$ could have been large enough to explain the preliminary (and later retired) results of the BICEP collaboration, which were pointing to $r\approx0.2$ \cite{Bicep}.

In critical Higgs inflation, $\xi$ is small enough so that problems related to the violation of unitarity (see e.g. \cite{Bezrukov:2010jz}) might be evaded, even though higher-dimensional operators might play an important role \cite{Bezrukov:2017dyv}. Recently, it has also been shown that this scenario is safe from the fine tuning associated to the initial conditions \cite{Salvio:2017oyf}. 

It has been recently proposed that critical Higgs inflation is a viable mechanism to produce primordial black holes constituting a fraction or a significant part of the dark matter observed today \cite{Ezquiaga:2017fvi}. 
This applies in general to potentials with an inflection point \cite{Garcia-Bellido:2017fdg, Ballesteros:2017fsr} or a local minimum \cite{Kannike:2017bxn} followed, at higher field values, by a plateau.
The results of ref. \cite{Ezquiaga:2017fvi} were questioned in \cite{Bezrukov:2017dyv}, where it 
was suggested that \cite{Ezquiaga:2017fvi} introduces a too large running of the non-minimal coupling, 
%(b_xi \simeq 20-40 depending on the version of that paper) 
which could not follow from the Standard Model (SM) non-minimally coupled to gravity\footnote{
This large running might appear in some SM extensions or maybe
through some non-perturbative physics, but those are scenarios that 
differ from the original Higgs inflation idea \cite{Bezrukov:2017dyv}.}.

Now that the tensor-to-scalar ratio is better constrained, $r<0.12$ at $95\%$ C.L. \cite{Ade:2015lrj, Ade:2015tva}, and in view of its possible applications in the phenomenology of primordial black holes, it is interesting to have a more robust understanding of the critical Higgs inflation scenario. The aim of this work is precisely to improve the robustness of the predictions of the cosmological observables (like $r$ and $n_s$), linking them to the present experimental range of the low energy parameters which control the shape of the Higgs effective potential - the strong coupling constant, $\alpha_s$, the top quark mass, $m_t$, and the Higgs boson mass, $m_H$ -, and assessing the size of the theoretical error associated to the calculation.

There is general agreement (see \cite{Iacobellis:2016eof} and references therein) on the fact that stability of the SM potential - and so the inflection point configuration - displays a tension with the low energy parameters at about $2\,\sigma$. More precisely, assuming the theoretical error associated to the Next-to-Next-to-Leading (NNLO) calculation to go in the "right direction",  stability requires for instance that $\alpha_s$, $m_t$ and $m_H$ take respectively their upper $2\,\sigma$, lower $1\,\sigma$ and central values \cite{Iacobellis:2016eof}. 

In the case of critical Higgs inflation, also the observational constraints on $r$ and $n_s$ have to be fulfilled.
The value of the SM Higgs potential at the inflection point, $V_i$, is particularly important for the prediction of $r$.
Once the low energy parameters are fixed, such value is subject to the theoretical errors associated to the various steps of the calculation: matching, running and effective potential expansion. These errors have been carefully studied in \cite{Iacobellis:2016eof}. For instance, it was shown that, even using the RGE-improved tree-level potential at NNLO, a large theoretical error plagues the value of $V_i$, so that one must consider at least the 1-loop effective potential to obtain a reliable result \cite{Iacobellis:2016eof}. The latter work focussed on the case of a rising inflection point in the SM, without any coupling to gravity (namely the possibility of a shallow false vacuum and its applications to cosmology \cite{MasinaHiggsmass,Masinatop,Masinahybrid,Masinaupgrade}). 
Here we extend the calculation by including the effect of  the non-minimal coupling to gravity. 

We find that the value of the Higgs potential at the inflection point is higher than what was considered in previous analyses \cite{Bezcritical, Hamada:2014iga, Hamada:2014wna, Bezrukov:2017dyv, Salvio:2017oyf},
and in particular it is much higher than the range required in \cite{Ezquiaga:2017fvi} for the issue of black holes. 
Fixing the amplitude of scalar perturbations at its observed value, it turns out that the prediction for $r$ is then accordingly higher.
We will show that,  even taking into account the NNLO theoretical error, the present upper bound on $r$ can be accommodated only at the price of assuming that $\alpha_s$ takes its upper $3-4\,\sigma$ value\footnote{ 
Notice that the Higgs false vacuum model was ruled out for precisely the same reason \cite{Iacobellis:2016eof}.}.

The model of critical Higgs inflation is thus in serious trouble {\it per se}, and it is quite unrealistic that it might account for a significant fraction of the dark matter seen today under the form of primordial black holes. 

The paper is organized as follows. In section 2, following \cite{Iacobellis:2016eof}, we review how to determine the Higgs potential in the SM according to the present state of the art, and discuss in particular the inflection point configuration. Section 3 is devoted to the model of Higgs inflation, while section 4 discusses the phenomenology of the inflection point configuration in the case of critical Higgs inflation. We draw our conclusions in section 5.

\vskip 1.cm
%%%%%%%%%%%%%%%%%%%%%%%%
\section{Higgs potential in the SM at NNLO}

Before introducing the model with the non-minimal coupling to gravity, we review the  findings of ref. \cite{Iacobellis:2016eof} about the rising inflection point configuration of the SM Higgs effective potential, as they will turn out to be relevant also in the case of the non-minimal coupling.

According to our conventions, the potential for the Higgs field $\phi$ contained in the Higgs doublet 
$\mathcal{H}=(0\quad(\phi+v)/\sqrt{2})^T$ is given, at tree-level, by
\begin{equation}
V(\phi)=\frac{\lambda}{6} \left(\left|\mathcal{H}\right|^2 - \frac{v^2}{2} \right)^2 \approx  \frac{\lambda}{24}  \phi^4\,,
\label{eq-Vtree}
\end{equation}
where $\lambda$ is the Higgs quartic coupling,  $v=1/(\sqrt2 G_\mu)^{1/2}=246.221{\rm ~GeV}$ and $G_\mu$ is the Fermi constant from muon decay \,\cite{Agashe:2014kda} and
the right hand side of eq.\,(\ref{eq-Vtree}) holds when considering large field values. 
Within our normalization, the mass of the Higgs boson and the mass of the fermion $f$ are given by the tree-level relations
\beq   
m_H^2= \frac{\lambda v^2} {3}\,\,\, , \,\,\, m_f =   \frac{h_f  v}{\sqrt{2}}\,\,\,, 
\eeq
where $h_f$ denotes the associated Yukawa coupling.

In order to extrapolate the behavior of the Higgs potential at very high energies, we adopt the $\overline{\rm MS}$ scheme 
and consider the matching and RGE evolution of the relevant couplings which, in addition to the Higgs quartic coupling $\lambda$, are: the three gauge couplings $g$, $g'$, $g_3$, the top Yukawa coupling $h_t$, and the anomalous dimension of the Higgs field $\gamma$. We then compute the   
RGE-improved Higgs effective potential at the Next-to-Next-to-Leading Order (NNLO), that is at the 2-loop level. 

Before discussing the procedure associated to matching, running, and effective potential expansion, we review the basic ideas of the RGE:
in applications where the effective potential $V_{\text{eff}}(\phi)$ at large $\phi$ is needed, as is the case for our analysis, 
potentially large logarithms appears, of the type $\log(\phi/\mu)$ where $\mu$ is the renormalization scale, which may spoil the applicability of perturbation theory.  The standard way to treat such logarithms is by means of the RGE.
The fact that, for fixed values of the bare parameters, the effective potential must be independent of the renormalization scale $\mu$, means that \cite{Coleman}
\beq
\left( \mu \frac{\partial}{\partial \mu} + \beta_i \frac{\partial}{\partial \lambda_i} - \gamma \frac{\partial}{\partial \phi}  \right) V_{\text{eff}}=0 \, ,
\eeq
where 
\beq
\beta_i = \mu \frac{d \lambda_i}{d \mu} \,\, , \,\, \gamma = -\frac{\mu}{\phi} \frac{d\phi}{d\mu}\,,
\eeq
are the $\beta$-functions corresponding to each of the SM couplings $\lambda_i$, and the anomalous dimension of the background field respectively.

The formal solution of the RGE is
\beq
V_{\text{eff}}(\mu,\lambda_i,\phi)= V_{\text{eff}}(\mu(t),\lambda_i(t),\phi(t))\,,
\label{eq-Veff}
\eeq
where 
\beq
\mu(t)=e^t \mu  \, ,\,\, \phi(t) =e^{\Gamma(t)} \phi \,,\,\, \Gamma(t)=- \int_0^t \gamma(\lambda(t')) dt' \, ,
\label{eq-mu}
\eeq
and $\lambda_i(t)$ are the SM running couplings, determined by the equation
\beq
\frac{d \lambda_i(t)}{dt} = \beta_i (\lambda_i(t)) \, ,
\eeq
and subject to the boundary conditions $\lambda_i(0)=\lambda_i$.
The usefulness of the RGE is that $t$ can be chosen in such a way that the convergence of
perturbation theory is improved, which is the case for instance when $\phi(t)/\mu(t) = {\cal{O}} (1)$.
In our calculation the boundary conditions are given at the top quark mass, $m_t$: we will then take $\mu=m_t$ in eq. (\ref{eq-mu}) from now on.

%%%
\subsection{Matching and running}
\label{sec-match}

In order to derive the values of the relevant parameters ($g$, $g'$, $g_3$, $h_t$, $\lambda$) at the top pole mass, $m_t$, we exploit the results of a detailed analysis about the matching procedure, performed by Bednyakov et al. \cite{Bednyakov:2015sca}. We refer the interested reader to \cite{Bednyakov:2015sca} and \cite{Iacobellis:2016eof} for more details; 
here we just mention our reference values:
\begin{itemize}
\item
for the strong coupling constant at $m_Z$, $\alpha_s^{(5)}$: we take the present \cite{Patrignani:2016xqp} world average experimental value of the strong coupling constant at $m_Z$, $\alpha_s^{(5,exp)}=0.1181$, and its associated $1\,\sigma$ error, $\Delta \alpha_s^{(5,exp)}=0.0013$;
\item
for the Higgs mass, $m_H$: we take the combined ATLAS and CMS result (after Run1) at $1\,\sigma$,
$m_H^{exp}=125.09$ GeV and $\Delta m_H^{exp} =0.24$ GeV \cite{ Aad:2015zhl};
\item 
for the top pole mass, $m_t$: we take the present combined Tevatron and LHC value of the MC top mass, 
$m_t^{MC}=(173.34 \pm 0.76$) GeV \cite{ATLAS:2014wva}; the uncertainty in the identification between the pole and MC top mass is currently estimated to be of order $200$ MeV \cite{Nason:2016tiy, Corcella:2015kth} (or even $1$ GeV for the most conservative groups \cite{Moch:2014tta}).
\end{itemize}

The $\beta$-functions can be organized as a sum of contributions with increasing number of loops:
\beq
 \frac{d}{d t} \lambda_i(t)=\kappa \beta_{\lambda_i}^{(1)}+\kappa^2  \beta_{\lambda_i}^{(2)} +\kappa^3  \beta_{\lambda_i}^{(3)}  + ...\, ,  \label{eq-RGE}
\eeq
where $\kappa = 1/(16 \pi^2)$ and the apex on the $\beta$-functions represents the loop order.
Here, we are interested in the RGE dependence of the couplings ($g$, $g'$, $g_3$, $h_t$, $\lambda$, $\gamma$). 
The 1-loop and 2-loop expressions for the $\beta$-functions in the SM are well known and can be found {\it e.g.} in Ford et al. \cite{Ford:1992}. The complete 3-loop $\beta$-functions for the SM have been computed more recently in refs. 
\cite{Mihaila:2012,Mihaila1, ChetyrkinZoller,Chetyrkin:2013, BednyakovPikelnerVelizhanin,BednyakovPikelnerVelizhanin1,Bednyakov:2013, Bednyakov:2014}.
The dominant 4-loop contribution to the running of the strong gauge coupling has been also computed recently, see refs. \cite{Zoller:2015tha,Bednyakov:2015}. In the present analysis we include all these contributions, as already done in ref. \cite{Iacobellis:2016eof}.

%%%%%%%%%%%%
\subsection{RGE-improved effective potential}\label{eff}

Without sticking to any specific choice of scale, the RGE-improved effective potential at high field values can be rewritten as
\beq
%V_{\text{eff}}(\mu(t),\lambda_i(t),\phi(t))
V_{\text{eff}}(\phi,t) 
\approx \frac{\lambda_{\text{eff}}(\phi,t)}{24} \phi^4\,,  
\eeq
where $\lambda_{\text{eff}} (\phi,t)$ takes into account the wave-function normalization and can be expanded as sum of tree-level plus increasing loop contributions: 
\beq
\lambda_{\text{eff}}(\phi,t)= e^{4 \Gamma(t)} 
\left[   \lambda(t) +  \lambda^{(1)}(\phi,t) +  \lambda^{(2)}(\phi,t) +... \right]\,.
\eeq

In particular, the 1-loop Coleman-Weinberg contribution \cite{Coleman:1973jx} is
\beq
\lambda^{(1)}(\phi,t)= 6 \frac{1}{(4 \pi)^2} \sum_p N_p \kappa^2_p(t) \left(  \log \frac{\kappa_p(t) e^{2 \Gamma(t)} \phi^2}{\mu(t)^2} -C_p \right)\,,    
\label{eqla1}
\eeq
where, generically, $p$ runs over the contributions of the top quark $t$, the gauge bosons $W$ and $Z$, the Higgs boson $\phi$ and the Goldstone bosons $\chi$.
The coefficients $N_p$, $C_p$, $\kappa_p$ are listed in table \ref{tab-coeff} for the Landau gauge (see {\it e.g.} table 2 of ref. \cite{DiLuzio:2014bua} for a general $R_\xi$ gauge). 

\begin{table}[h!]
\vskip .5cm 
\centering

\begin{tabular}{  r | c  c  c   c  c      }
 $p$ & $t$ & $W$ & $Z$ & $\phi$  & $\chi$  \\
\hline
$N_p$         & $-12$ & $6$ & $3$  & $1$ & $3$   \\
$C_p$         &  $3/2$ & $5/6$  & $5/6$ & $3/2$ & $3/2$  \\
$\kappa_p$ &  $h^2/2$ & $g^2/4$ & $(g^2+g'^2)/4$ & $3\lambda$ & $\lambda$
\end{tabular}
\caption{Coefficients for eq.\,(\ref{eqla1}) in the Landau gauge. }
\label{tab-coeff}
\vskip .5cm 
\end{table}

The 2-loop contribution $\lambda^{(2)}(\phi, t)$ was derived by Ford et al. in ref. \cite{Ford:1992} and, in the limit $\lambda\rightarrow0$,  was cast in a more compact form in refs. \cite{Degrassi,Buttazzo:2013uya}. 
We verified, consistently with these works, that the error committed in this approximation is less than $10\%$ and can thus be neglected.

It is clear that when $\lambda(t)$ becomes negative, the Higgs and Goldstone contributions in eq.\,(\ref{eqla1}) are small but complex, and this represents a problem in the numerical analysis of the stability of the electroweak vacuum. Indeed, 
in refs. \cite{Degrassi,Buttazzo:2013uya} the potential was calculated at the 2-loop level, but setting to zero the Higgs and Goldstone contributions in eq.\,(\ref{eqla1}). 
Some authors \cite{Elias-Miro:2014pca, Andreassen:2014gha} recently showed that the procedure of refs. \cite{Degrassi,Buttazzo:2013uya} is actually theoretically justified when $\lambda$ is small (say $\lambda \sim \hbar$): in this case, the sum over $p$ does not have to include the Higgs and Goldstone's contributions, which rather have to be accounted for in the 2-loop effective potential,
which practically coincides with the expression derived in refs. \cite{Degrassi,Buttazzo:2013uya}. 
For the rising inflection point configuration we are interested in, $\lambda$ is indeed small: as already done in ref. \cite{Iacobellis:2016eof}, we thus adopt the procedure outlined in \cite{Andreassen:2014gha}. Here, however, we prefer to work with the wave-function renormalized field, $\phi(t)$, instead of the classical one, $\phi$. 
Explicitly:
\beq
V_{\rm eff} = V^{(0)} +V^{(1)}  +V^{(2)} +....
\eeq
where
\beq
V^{(0)} = \frac{\lambda(t)}{24}  \phi(t)^4\,\,\,,
\eeq
\bea
V^{(1)}&=& \frac{1}{24} \frac{6}{(4 \pi)^2} 
 \left[  
6   \left(\frac{g(t)^2}{4} \right)^2 \left(  \log \frac{ \frac{g(t)^2}{4} \phi(t)^2}{\mu(t)^2} -\frac{5}{6} \right) \right. \nonumber \\
&+& 3 \left(\frac{g(t)^2+{g'(t)}^2}{4} \right)^2 \left(  \log \frac{ \frac{g(t)^2+{g'(t)}^2}{4}  \phi(t)^2}{\mu(t)^2} - \frac{5}{6} \right)    \nonumber \\
&-&  \left. 12 \left( \frac{h(t)^2}{2}  \right)^2 \left(  \log \frac{\frac{h(t)^2}{2}  \phi(t)^2}{\mu(t)^2} -\frac{3}{2} \right) 
 \right]   \phi(t)^4 \,,    
\label{eqV1}
\eea
and $V^{(2)}$ can be found in \cite{Degrassi,Buttazzo:2013uya}.

A relevant aspect of the present calculation is represented by the well-known fact that the RGE-improved effective potential is gauge dependent.
After choosing the renormalization scale $t$, the RGE-improved effective potential,
$V_{\text{eff}}( \phi, \xi)$, is a function of $\phi$,  the gauge-fixing parameters collectively 
denoted by $\xi$, and the other input parameters as $m_t$, $m_H$, $\alpha_s^{(5)}$.
Due to the explicit presence of $\xi$ in the vacuum stability and/or inflection point conditions, 
it is not obvious a priori which are the physical (gauge-independent) observables entering the vacuum
stability and/or inflection point analysis. 
The basic tool, in order to capture the gauge-invariant content of the effective
potential is given by the Nielsen identity \cite{Nielsen:1975fs}
\beq
\left(\xi   \frac{\partial }{\partial \xi}  + C(\phi,\xi)   \frac{\partial  }{\partial \phi}  \right)V_{\text{eff}}(\phi,\xi)=0\,,
\eeq
where $C(\phi, \xi)$ is a correlator whose
explicit expression will not be needed for our argument.
The equation means that $V_{\text{eff}}(\phi,\xi)$ is constant along the characteristics of the equation,
which are the curves in the $(\phi, \xi)$ plane for which $d\xi= \xi /C(\phi,\xi) d\phi $.
In particular, the identity says that the effective potential is gauge independent where it is stationary, as happens for two degenerate vacua and for the inflection point configuration.
One can also show \cite{Iacobellis:2016eof} that the peculiar values of the low energy input parameters (as for instance $m_t$, the Higgs mass and $\alpha_s^{(5)}$) ensuring stationary configurations
are gauge independent.

Working in the Landau gauge is thus perfectly consistent in order to calculate the value  of the effective potential at a stationary point, call it $V_s$, or the value of the input parameters providing it. 
Nevertheless, one has to be aware that the truncation of the effective potential loop expansion at some loop order, introduces an unavoidable theoretical error both in $V_s$ and in the input parameters. %$m^s_t$. 
For this sake, it is useful to define the parameter $\alpha$ via
\beq
\mu(t)= \alpha \, \phi(t) \, ,
\label{eq-alfa}
\eeq
and study the dependence of $V_s$ and the input parameters 
on $\alpha$. The higher the order of the loop expansion to be considered, the less the dependence on $\alpha$. This was shown explicitly in \cite{Iacobellis:2016eof}, where we studied the case of two degenerate vacua and the case of a rising inflection point, respectively\footnote{Actually, in \cite{Iacobellis:2016eof} we defined $\alpha$ in a slightly different way than we do here, namely
in terms of the classical field $\phi$, rather than the wave-function renormalized one, $\phi(t)$. There is no conceptual difference in doing so, and the numerical difference in $\alpha$ is marginal.}. In the following we summarize and elaborate on the main results, as they will be useful also for the analysis of critical Higgs inflation.

%%%%%
\subsection{Two degenerate vacua}
\label{sec-two}

As discussed in the previous section, once $m_H$ and $\alpha_s^{(5)}$ have been fixed, 
the value of the top mass for which the SM displays two degenerate vacua, $m^c_t$, is a gauge invariant quantity.
This value is however plagued by experimental and theoretical errors.
The result of the NNLO calculation is \cite{Iacobellis:2016eof}:
\beq
m^c_t = (171.08 \pm 0.37_{\alpha_s} \pm 0.12_{m_H}  \pm 0.32_{th} )\, {\rm GeV}\,,
\label{eq-mct}
\eeq
where the first two errors are the $1\,\sigma$ variations of $\alpha_s^{(5)}$ and $m_H$.
Our results for the value of $m^c_t$ update and improve
but, modulo the doubling of the experimental error in $\alpha_s^{(5)}$, are essentially consistent with those of the literature \cite{Degrassi, Bezrukov:2012sa, Alekhin:2012py, Masina:2012tz, Buttazzo:2013uya,Bednyakov:2015sca}. 

\begin{figure}[t!]
%\vskip 1.5cm 
 \begin{center}
\includegraphics[width=9cm]{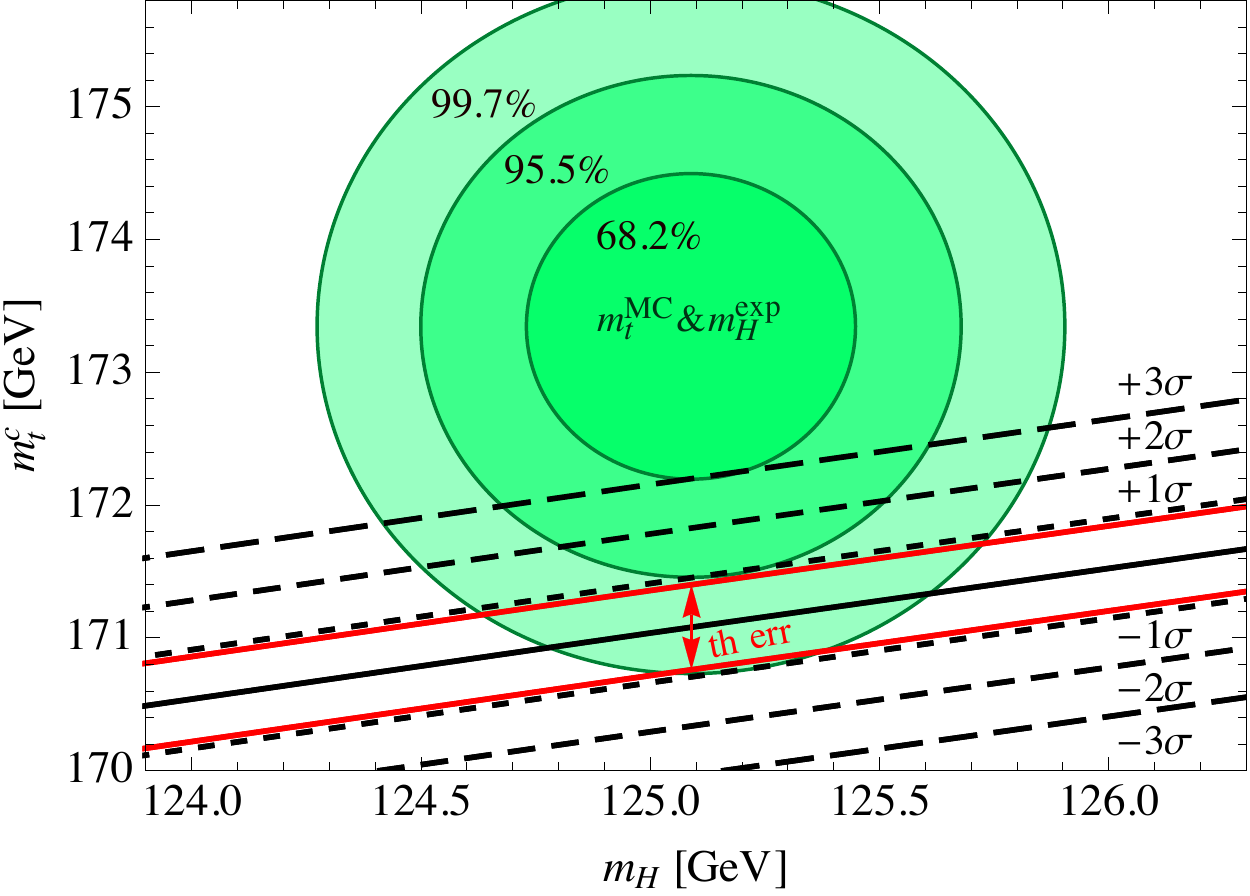}   
 \end{center}
\caption{\baselineskip=15 pt \small  
Lines for which the Higgs potential develops a second degenerate minimum at high energy. The solid line corresponds to the central value of $\alpha_s^{(5)}$; the dashed lines are obtained by varying $\alpha_s^{(5)}$ in its experimental range, up to $3\,\sigma$. The (red) arrow represents the theoretical error in the position of the lines. The (green) shaded regions are the covariance ellipses obtained combining $m_t^{MC}=173.34 \pm 0.76$ GeV and  $m_H^{exp}=125.09\pm 0.24 $ GeV; the probability of finding $m_t^{MC}$ and $m_H$ inside the inner (central, outer) ellipse is equal to $68.2\%$ ($95.5\%$, $99.7\%$). The plot is taken from ref. \cite{Iacobellis:2016eof}.}   
\label{fig-mtmH}
\vskip .5 cm
\end{figure}

In fig.\,\ref{fig-mtmH}, $m_t^c$ is displayed as a function of $m_H$ for selected values of $\alpha_s^{(5)}$; in particular, the solid line refers to its central value, while the dotted, short and long dashed lines refer to the $1\,\sigma$, $2\,\sigma$ and $3\,\sigma$ deviations respectively. In the region below (above) the line the potential is stable (metastable). The theoretical uncertainty on $m_t^c$ due to the NNLO matching turns out to be about $\pm 0.32$ GeV: the position of the straight lines in fig.\,\ref{fig-mtmH} can be shifted up or down, as represented by the (red) arrow for the central value of $\alpha_s^{(5)}$.
The value $\pm 0.32$ GeV is obtained combining in quadrature the error on $m_t^c$ associated to the matching of $\lambda$, $\pm 0.19$ GeV for $\pm \Delta \lambda$, and the one associated to the matching of the top Yukawa coupling, $\mp 0.25$ GeV for $\pm \Delta y_t$. 

The present combined Tevatron and LHC value of the MC top mass is $m_t^{MC}=(173.34 \pm 0.76$) GeV \cite{ATLAS:2014wva}.
Taking into account the theoretical error, we see that the stability line for the central (upper $2\,\sigma$) value of $\alpha^{(5)}_s$ touches the $m^{MC}_t-m_H^{exp}$ covariance ellipse corresponding to a $95.5\%$ ($68.2\%$) probability.
This calculation of the experimental and theoretical uncertainties on $m_t^c$, in addition to the uncertainty in the identification of the MC and pole top masses, lead us to conclude that stability is at present still compatible with the experimental data at about $2\,\sigma$ \cite{Iacobellis:2016eof}.

%%%%%%
\subsection{Rising inflection point}
\label{sec-inf}

Such configuration is relevant for the class of models of primordial inflation based on a shallow false minimum \cite{MasinaHiggsmass,Masinatop,Masinahybrid,Masinaupgrade}, which was studied in \cite{Iacobellis:2016eof}, and those based on the non-minimal coupling, which we study in the present work.

The value of the top mass giving the inflection point configuration, $m^i_t$, is smaller but so close to the one giving two degenerate vacua that eq. (\ref{eq-mct}) applies also in this case. 

\begin{figure}[t!]
%\vskip .5cm 
 \begin{center}
  \includegraphics[width=8cm]{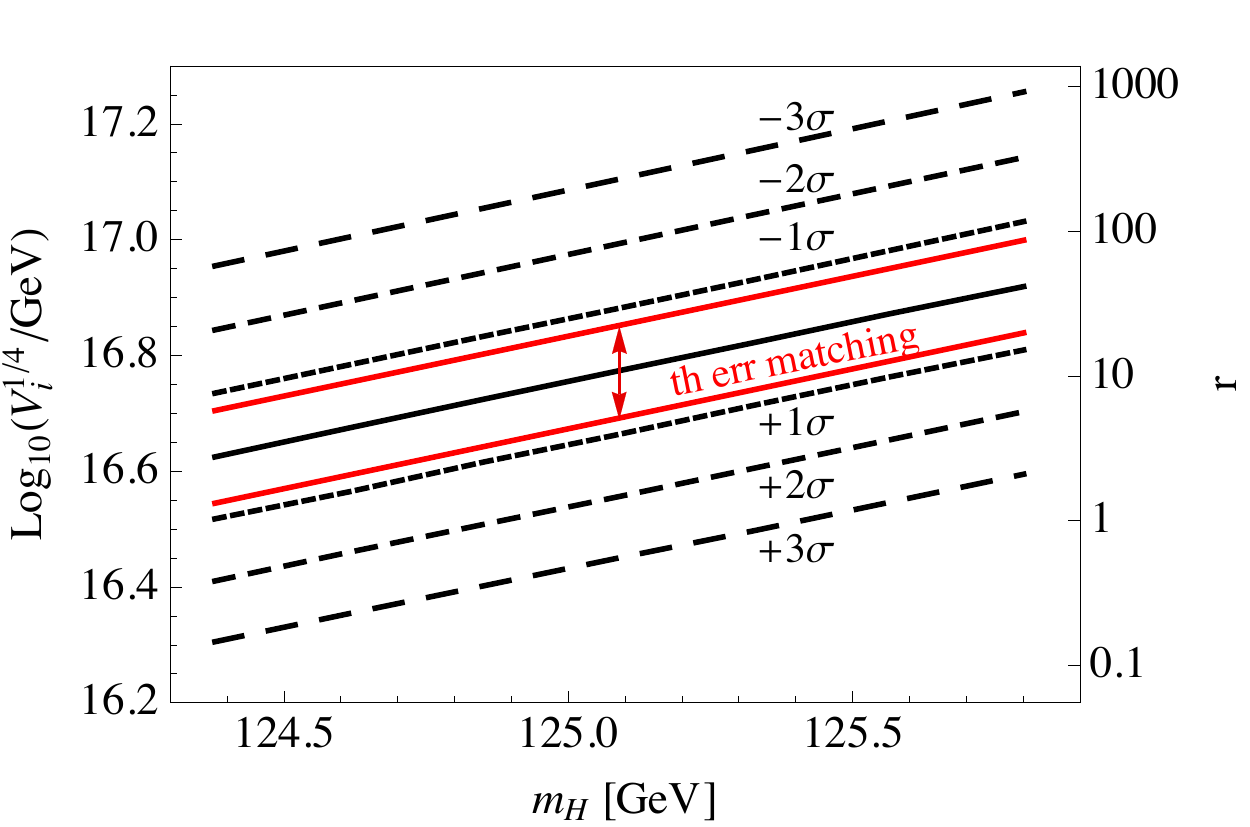}   \,\,\,\,\,\,
  \includegraphics[width=7.2cm]{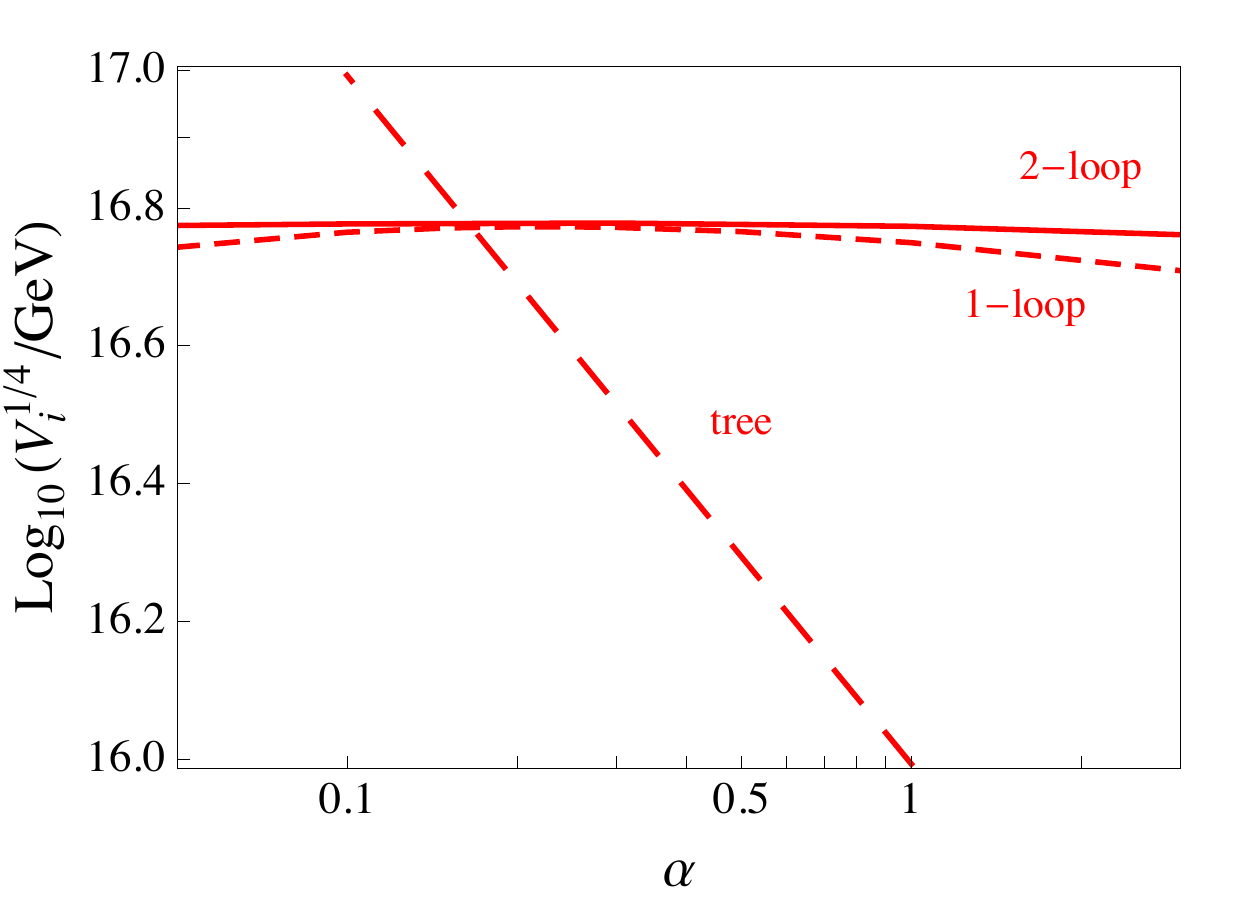}   
 \end{center}
\caption{\baselineskip=15 pt \small 
Left: Dependence of $V_i^{1/4}$ on $m_H$ for fixed values of $\alpha_s^{(5)}$. The (red) arrow and solid lines show the theoretical error due to the matching of $\lambda$. The right vertical axis displays the associated value of the tensor-to-scalar ratio $r$, according to eq.\,(\ref{eq-r}). The plot is taken from ref. \cite{Iacobellis:2016eof}. Right: Dependence of $V_{i}^{1/4}$ on $\alpha$ at the tree, 1-loop and 2-loop levels. For definiteness, $\alpha_s^{(5)}$ and $m_H$ are assigned to their central values. } 
\label{fig-Veff}
\vskip .1 cm
\end{figure}

We denote the value of the Higgs effective potential at the inflection point by $V_i$.
Experimental uncertainties on $V_i$ can be estimated as follows: we let $\alpha_s^{(5)}$ vary in its $3\,\sigma$ experimental range and, for fixed values of $m_H$, we determine $m^i_t$ and the 2-loop effective potential $V_i$:
the result is displayed in the left panel of fig.\,\ref{fig-Veff} (taken from \cite{Iacobellis:2016eof}): one can see that increasing $\alpha_s^{(5)}$
from its lower to its upper $3\,\sigma$ range, $ V_i^{1/4}$ decreases from $2\times 10^{17}$ GeV up to $2 \times 10^{16}$ GeV; the dependence on $m_H$ is less dramatic.

Theoretical errors can be divided in three categories: those associated to i) the matching, ii) the running, and iii) the effective potential expansion.

i) Theoretical errors associated to the NNLO matching of $\lambda$  are displayed via the (red) lines in the left panel of fig.\,\ref{fig-Veff}: the line associated the central value of $\alpha_s^{(5)}$ could be shifted by about $\pm 0.08$ when the quartic coupling changes by $\pm \Delta \lambda$. This theoretical error is thus slightly smaller than the experimental error due to the $1\,\sigma$ variation of $\alpha_s^{(5)}$. The theoretical error in the matching of the top Yukawa coupling has a negligible effect on $V_i$.

ii) The order of magnitude of the theoretical errors associated to the $\beta$-functions at NNLO can be estimated by studying the impact of the subsequent correction; it turns out that such error is negligible.

iii) The theoretical uncertainty associated to the fact that we truncate the effective potential at some loop level can be estimated by studying the dependence of $V_{i}$ on the parameter $\alpha$ defined via eq. (\ref{eq-alfa}). We fix $\alpha_s^{(5)}$ and $m_H$ at their central values and display in the right panel of fig.\,\ref{fig-Veff} the resulting value of $V_{i}^{1/4}$ at the tree, 1-loop and 2-loops levels by means of the long-dashed, dashed and solid lines respectively.
The dependence of $V_{i}^{1/4}$ on $\alpha$ at the tree-level is implicit, 
$V_{\text{eff}} \propto \lambda\left( \ln( \alpha  \phi(t) / m_t ) \right)$, but significant: it is uncertain by one order of magnitude when $\alpha$ is varies in the interval $0.1-1$.
The 1-loop corrections flattens the dependence on $\alpha$ so that, in the interval $0.1-1$, the uncertainty on $V_i^{1/4}$ gets reduced down to about $5\%$, much smaller that the theoretical one due to the matching;
the 2-loop correction further flattens the dependence on $\alpha$ and allows to estimate $V_{i}^{1/4}$ with a $1\%$ precision.

Summarizing, the result of the NNLO calculation is \cite{Iacobellis:2016eof}:
\beq
\log_{10} ({V}_{i}^{1/4} /{\rm GeV})=  16.77  \pm 0.11_{\, \alpha_s} \pm 0.05_{m_H}  \pm 0.08_{th}  \,,
\label{eq-V14}
\eeq
where the first two errors refer to the $1\,\sigma$ variations of $\alpha_s^{(5)}$ and $m_H$ respectively, 
while the theoretical error is dominated by the one in the matching of $\lambda$.

%%%%
\subsubsection{Impact on models of inflation with a rising inflection point}

A precise determination of $V_i$ is important for models of inflation based on the idea of a shallow false minimum \cite{MasinaHiggsmass,Masinatop,Masinahybrid,Masinaupgrade} as, in these models, $V_i$ and the ratio of the scalar-to-tensor modes of primordial perturbations, $r$, are linked via:
\beq
V_{i}  = \frac{3 \pi^2}{2} \, r \, A_s \, ,
\label{eq-r}
\eeq
where $A_s =2.2 \times 10^{-9} $ \cite{Ade:2015xua} is the amplitude of scalar perturbations.
This relation follows from the fact that about $62$ e-folds before the end of inflation, the Higgs field (playing the role of a curvaton) is at the inflection point, so that
\beq
A_s= %P_s(N=62) \simeq 
\frac{H^2}{8 \pi^2 \epsilon} \left.  \right|_{N=62} 
%\simeq  \frac{V_i}{24 \pi^2 \epsilon}\left.   \right|_{N=62} 
%\simeq  \frac{2V_i}{3 \pi^2 r} %\left.   \right|_{N=62} 
\,\, ,
\eeq
where $H^2 \approx V_i/3 $ is the Hubble parameter (dominated by the SM potential), and the inflaton is in a slow-roll phase, so that $r= 16 \epsilon$.

In view of such application, the right axis of the plot in the left panel of fig.\,\ref{fig-Veff} reports the corresponding value of $r$. The dependence of $r$ on $\alpha_s^{(5)}$ is strong: when the latter is varied in its $3\,\sigma$ range, $r$ spans about three orders of magnitude, from $0.3$ to $300$. The dependence on $m_H$ is milder.
The theoretical error in the matching of $\lambda$ implies an uncertainty on $r$ by a factor of about $2$.

\begin{figure}[t!]
%\vskip .5cm 
 \begin{center}
  \includegraphics[width=9cm]{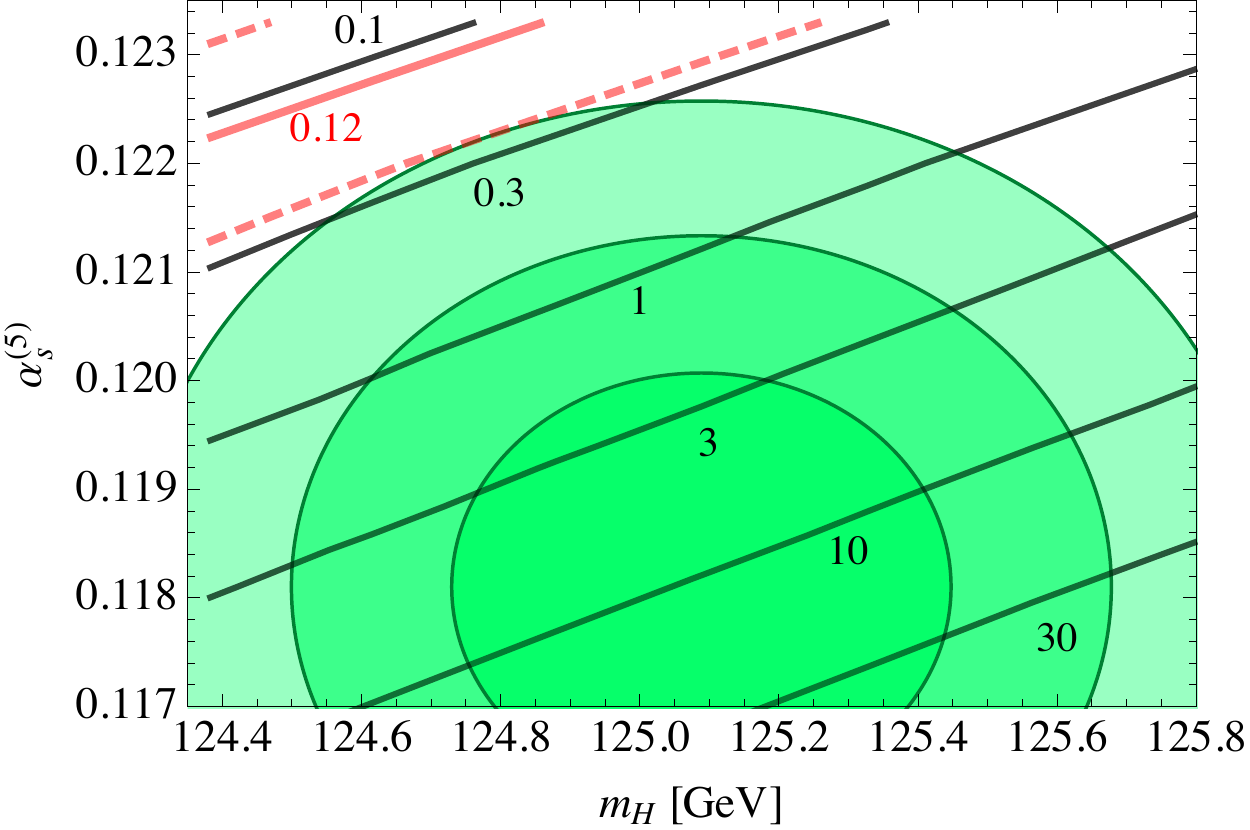}   
 \end{center}
\caption{\baselineskip=15 pt \small Contour levels of $r$ in the plane $(m_H, \alpha_s^{(5)})$. The theoretical uncertainty corresponding to $r=0.12$ is shown by means of the (red) dashed lines. The shaded regions are the covariance ellipses indicating that the probability of finding the experimental values of $m_H$ and $ \alpha_s^{(5)}$  inside the ellipses are respectively $68.2\%, 95.4\%,99.7\% $. The plot is taken from ref. \cite{Iacobellis:2016eof}. }
\label{fig-r}
\vskip .1 cm
\end{figure}

According to the 2015 analysis of the Planck Collaboration, the present upper bound on $r$ at the pivot scale $k_*=0.002$ Mpc$^{-1}$ is  $r <0.12$ at $95\%$ C.L. \cite{Ade:2015lrj}, as also confirmed by the recent joint analysis with the BICEP2 Collaboration \cite{Ade:2015tva}. 
Due to eq.\,(\ref{eq-r}), this would translate into the $95\%$ C.L. bound 
\beq
\log_{10} ({ V}_i^{1/4}/{\rm GeV} )< 16.28\,\, ,
\eeq
%$\bar V_{i}^{1/4} <  1.9\times 10^{16}$ GeV 
which implies a tension with eq.\,(\ref{eq-V14}) at about $4\,\sigma$ with respect to $\alpha_s^{(5)}$.
This tension might be reduced at about $3\,\sigma$ assuming the theoretical error on the matching of $\lambda$ to go in the "right" direction of lowering $V_i$ (this would correspond to $+\Delta \lambda$, which however goes in the "wrong" direction for the sake of $m_t^c$). 
This can be graphically seen in fig.\,\ref{fig-r}, where the contour levels of $r$ in the plane $(m_H, \alpha_s^{(5)})$ are shown.
Even invoking the uncertainty due to the matching (lower red-dashed lines), a value for $r$ as small as $0.12$ (red-solid line), could be obtained only with $\alpha_s^{(5)}$ to take its upper $3\,\sigma$ value, and $m_H$ its lower $1\,\sigma$ one; 
the value of $m_t$ could stay around its lower $1.5\,\sigma$ value, as can be see from fig. \ref{fig-mtmH}.

These considerations will be useful also for the model of critical Higgs inflation, which however requires a specific study.  
In the next section we review this model, paying attention to work at least at 1-loop in the expansion of the effective potential.

\vskip 1.cm
%%%%%%%%%%%%%%%%%%%%%%%%
\section{Higgs Inflation: the model}

We introduce a non-minimal gravitational coupling $\xi$ between the SM Higgs doublet $\mathcal{H}$ and the Ricci scalar $R$\,\cite{BezHiggs}. The classical action for Higgs inflation is:
\begin{equation}\label{act}
\mathcal{S}=\int d^{4}x \, \sqrt{-g}\left[\mathcal{L}_{SM}-\frac{M_{P}^{2}}{2} R -\xi\left|\mathcal{H}\right|^{2} R\right]\,,
\end{equation}
where $\mathcal{L}_{SM}$ is the Standard Model Lagrangian, $M_{P}=1/(8\pi G_{N})^{1/2}\simeq2.43\times10^{18} \,{\rm GeV}$  is the reduced Planck mass and $g$ is the determinant of the Friedmann-Lem\^aitre-Robertson-Walker metric. 
The relevant part of the action\,(\ref{act}) from a cosmological point of view is:
\begin{equation}
\mathcal{S}_{J}=\int d^{4}x \, \sqrt{-g}\left[\left|\partial\mathcal{H}\right|^{2}-\frac{M_{P}^{2}}{2} R-\xi\left|\mathcal{H}\right|^{2} R-V\right]\,,
\end{equation}
where $V$ is the SM potential of eq. (\ref{eq-Vtree}), $\left|\partial\mathcal{H}\right|^{2}=(\partial_{\mu}\mathcal{H})^{\dagger}(\partial^{\mu}\mathcal{H})$ and the subscript $J$ means that the action is evaluated in the Jordan frame
(where physical distances are measured and the inflationary model is defined).
In order to remove the non-minimal coupling we introduce a conformal (or Weyl) transformation:
\begin{equation}\label{weyl}
%g_{\mu\nu}\rightarrow
\tilde{g}_{\mu\nu}=\Omega^{2}g_{\mu\nu},\qquad \Omega^{2}\equiv1+2\xi\frac{\left|\mathcal{H}\right|^{2}}{M_{P}^{2}}\,.
\end{equation}
If we further consider the unitary gauge, in which the only scalar field is the radial mode $\phi=\sqrt{2 \left|\mathcal{H}\right|^2 }$,
we obtain the Einstein frame action where gravity is canonically normalized:
\begin{equation}\label{acteinsim}
\mathcal{S}_{E}=\int d^{4}x\, \sqrt{-\tilde{g}}\left[-\frac{M_{P}^{2}}{2}\tilde{R}+K\frac{(\partial\phi)^{2}}{2}-\frac{V}{\Omega^{4}}\right]\,\,\,,\,\,
K = \frac{ \Omega^{2}+ \frac{3}{2} \left(   \frac{d\Omega^2}{d(\phi/M_P)} \right)^2  }{\Omega^{4}}  \,\, .
\end{equation}

From now on, the bar over a quantity will indicate that it is given in (reduced) Planck units. The kinetic term for the classical Higgs field $\phi$ in\,(\ref{acteinsim}) can be made canonical by the redefinition 
$\bar \phi= \bar \phi(\bar \chi)$:
\begin{equation}\label{redef}
\frac{d \bar \chi}{d\bar \phi}= \sqrt{K}=
\frac{\sqrt{ 1+\xi \bar \phi^2 + 6 (\frac{1}{2} \frac{d\xi}{d\bar\phi} \bar \phi+ \xi)^2   \bar \phi^2} } { 1+ \xi  \bar\phi^2}
%\sqrt{\frac{\Omega^{2}+  \frac{3}{2} \left(   \frac{d\Omega^2}{d\phi} \right)^2 }{\Omega^{4}}}
\,,\qquad\bar \chi(\bar \phi=0)=0\,\,.
\end{equation}

%Since $\phi(\chi)$ is invertible, one can obtain a closed analytical relation between the two fields:
%\beq
%\bar \chi( \bar \phi)= \sqrt{\frac{1+6\xi}{\xi}}\sinh^{-1}\left( \sqrt{\xi(1+6\xi)} \bar \phi \right) 
%-\sqrt{6} \tanh^{-1}\left(\frac{\sqrt{6}\xi \bar \phi}{\sqrt{ 1 +\xi(1+6\xi) \bar \phi^{2}}}\right)    \,.
%\label{eq-closed}
%\eeq

The final expression for the Einstein frame action is
\begin{equation}
\mathcal{S}_{E}=\int d^{4}x \, \sqrt{-\tilde{g}}\left[-\frac{M_{P}^{2}}{2}\tilde{R}+ \frac{(\partial\chi)^{2}}{2}-U\right]\,,
\end{equation}
where the potential $U$ felt by $\chi$ is 
\begin{equation}\label{utree}
U = \frac{V}{\Omega^{4}} \,\,.
\end{equation}
Hence, at tree-level
\begin{equation}\label{utree}
U =\frac{\lambda}{24}\frac{(\phi(\chi)^{2}-v^{2})^{2}}{(1+\xi\bar{\phi}(\chi)^{2})^{2}}\, 
\simeq \frac{\lambda}{24}\frac{\phi(\chi)^{4}}{(1+\xi\bar{\phi}(\chi)^{2})^{2}}\,\,.
\end{equation}
The potential is flat for large field values, $\bar \phi > 1/\sqrt{\xi} $, and can in principle provide a slow-roll inflationary phase.

%%%%%%%%%%
\subsection{Radiative corrections}

We turn to consider the inclusion of radiative corrections: the running of the couplings, now including also the running of the non-minimal coupling $\xi$, %see e.g. \cite{Allison, George:2015nza}, 
and the loop corrections to the effective potential. %, as done in \cite{Fumagalli:2016lls}. 

The expressions for the $\beta$-functions of the relevant SM couplings, including $\xi(t)$, 
can be found e.g. in refs. \cite{Allison, George:2015nza}. The running of $\xi(t)$ is not dramatic: going from $t=0$ (low energies) to $t_P= \ln(M_P/m_t)$ (Planck scales), it increases by about $15\%$.
The non-minimal coupling affects the running through the appearance of a factor $s$ that suppresses the contribution of the physical Higgs to the RGEs \cite{DeSimone:2008ei, Allison}:
\begin{equation}
s(\phi(t))=\frac{1+\xi(t) {\bar \phi(t)^{2}}}{1+(1+6\xi(t)) \xi(t) {\bar \phi(t)^{2}}}\,,
\end{equation}
where $\phi(t) = e^{\Gamma(t)} \phi$ is the wave-function renormalized field.
For small field values $\bar \phi(t) \ll 1/\sqrt{\xi(t)}$, $s\simeq1$, recovering the SM case; in the inflationary regime 
$\bar \phi(t)\gg 1/\sqrt{\xi(t)}$,  the RG equations differ from those of the SM as quantum loops involving the Higgs field are suppressed by $s\simeq 1/(1+6\xi(t))$.

The total RGE-improved effective potential is given by  
\beq
{U}_{\text{eff}}={U}^{(0)}+{U}^{(1)}+{U}^{(2)}+... \,,
\eeq
 with the running of all the couplings involved, evaluated at some renormalization scale $\mu(t)$, conveniently chosen in order to minimize the effect of the logarithms. 
There exist two options for the quantization of the classical theory  - see e.g. \cite{Allison, Fumagalli:2016lls} for recent reviews.
One can compute quantum corrections to the potential after the transformation\,(\ref{weyl}), in the Einstein frame (prescription I)\,\cite{BezHiggs} or before, directly in the Jordan frame (prescription II)\,\cite{Barvinsky}.

According to prescription I, the tree-level RGE-improved potential is first rewritten in the Einstein frame giving 
\begin{equation}
{U}^{(0)}= \frac{\lambda(t)}{24}  \frac{ { \phi(t)}^{4}} {\Omega(t)^4} \,\,\,, \,\,\Omega(t)^2=1+\xi(t) \, {\bar \phi(t)}^{2}\,\,.
\label{treeP}
\end{equation}
%where $\Gamma(t)$ takes into account the wave function renormalization of $\phi$.
The 1-loop corrections take the form of \,(\ref{eqV1}), but the particle masses are computed from the tree-level potential above: this means that the quantity $\kappa_p$ of table 1 displays a suppression factor $\Omega(t)^{2}$ for the $W,Z,t$ contributions (while the Higgs and Goldstone contribution belong to $U^{(2)}$):  
\bea
U^{(1)}&=& \frac{1}{24} \frac{6}{(4 \pi)^2} 
 \left[  
6   \left(\frac{g(t)^2}{4} \right)^2 \left(  \log \frac{ \frac{g(t)^2}{4}  \phi(t)^2}{\mu(t)^2 \Omega(t)^2} -\frac{5}{6} \right) \right. \nonumber \\
&+& 3 \left(\frac{g(t)^2+{g'(t)}^2}{4} \right)^2 \left(  \log \frac{ \frac{g(t)^2+{g'(t)}^2}{4}   \phi(t)^2}{\mu(t)^2 \Omega(t)^2} - \frac{5}{6} \right)    \nonumber \\
&-&  \left. 12 \left( \frac{h(t)^2}{2}  \right)^2 \left(  \log \frac{\frac{h(t)^2}{2}  \phi(t)^2}{\mu(t)^2 \Omega(t)^2} -\frac{3}{2} \right) 
 \right] \frac{ \phi(t)^4}{ \Omega(t)^4 } \,.    
\label{eqUa}
\eea
The 2-loop radiative corrections ${U}^{(2)}$ can be found in the same way, operating on the explicit form given in 
\cite{Degrassi,Buttazzo:2013uya}.
The appropriate scale for minimizing the effect of the logarithms is given by $\phi(t)/\Omega(t)$. Following the argument illustrated in the previous section, we define the parameter $\alpha$ via
\begin{equation}\label{renorm1}
\mu(t)=\alpha \frac{\phi(t)}{\Omega(t)}
\,. 
\end{equation}

According to prescription II, the radiative corrections are evaluated directly in the Jordan frame, before the conformal transformation: they are thus given by $V^{(1)}$ of eq.\,(\ref{eqV1}).
After going in the Einstein frame, the tree-level potential is thus the same as (\ref{treeP}), while $U^{(1)} =V^{(1)}/\Omega(t)^4$
becomes
\bea
U^{(1)}&=& \frac{1}{24} \frac{6}{(4 \pi)^2} 
 \left[  
6   \left(\frac{g(t)^2}{4} \right)^2 \left(  \log \frac{ \frac{g(t)^2}{4}  \phi(t)^2}{\mu(t)^2 } -\frac{5}{6} \right) \right. \nonumber \\
&+& 3 \left(\frac{g(t)^2+{g'(t)}^2}{4} \right)^2 \left(  \log \frac{ \frac{g(t)^2+{g'(t)}^2}{4}   \phi(t)^2}{\mu(t)^2 } - \frac{5}{6} \right)    \nonumber \\
&-&  \left. 12 \left( \frac{h(t)^2}{2}  \right)^2 \left(  \log \frac{\frac{h(t)^2}{2}  \phi(t)^2}{\mu(t)^2 } -\frac{3}{2} \right) 
 \right] \frac{  \phi(t)^4}{ \Omega(t)^4 } \,.
\label{eqU1b}
\eea
Now it make sense to define the parameter $\alpha$ precisely as in eq. (\ref{eq-alfa}), namely
\begin{equation}
\label{renorm2}
\mu(t)=\alpha\, \phi(t)\,\,.
\end{equation}

We can recognize that, due to the different choices of $\mu(t)$, the two prescriptions are formally equivalent up to 1-loop\footnote{This is due to the fact that for $\lambda$ small, the contribution of the Higgs and would de Goldstone bosons have to be included in the 2-loop contribution.}, as they both give:
\bea
U^{(0)}+U^{(1)}&=& \frac{1}{24}  \left(   \lambda(t)     +  \frac{6}{(4 \pi)^2} 
 \left[  
6   \left(\frac{g(t)^2}{4} \right)^2 \left(  \log \frac{ g(t)^2}{4 \alpha^2}   -\frac{5}{6} \right) \right.  \right. \nonumber \\
&+& 3 \left(\frac{g(t)^2+{g'(t)}^2}{4} \right)^2 \left(  \log  \frac{g(t)^2+{g'(t)}^2}{4 \alpha^2}   - \frac{5}{6} \right)    \nonumber \\
&-&  \left.  \left.12 \left( \frac{h(t)^2}{2}  \right)^2 \left(  \log \frac{h(t)^2}{2 \alpha^2}  -\frac{3}{2} \right) 
 \right] \right) \frac{  \phi(t)^4}{ (1+\xi(t) \, {\bar \phi(t)}^{2})^2 } \,.
\label{eqU1c}
\eea

So, in practice, the difference between the effective potentials at 1-loop for the two renormalization prescriptions is the relation between $t$ (the argument of the running couplings) and the wave-function renormalized field $\phi(t)$. 
For small field values, $\bar \phi(t) \ll 1/\sqrt{\xi(t)}$, the relation is the same for the two prescriptions, as in this case $\Omega \approx 1$. 
In the inflationary region where $\bar \phi(t) \gg 1/\sqrt{\xi(t)}\approx 1/\sqrt{\xi(t_P)}$, the situation changes:
for prescription I,  $t$ approaches a nearly constant value, approximately given by $\ln [\alpha/(\sqrt{\xi(t_P)} m_t)]$, 
and hence so do the couplings $g(t), g'(t),$ etc.;
for prescription II, $t=\ln [(\alpha \phi(t))/m_t]$ does not approach a constant value. 
As a result, the effective potential for prescription I approaches a constant value in the inflationary region (even after including radiative corrections) while the effective potential for prescription II, due to the continued running of the couplings, does not. 

This difference can have an impact on Higgs inflation and its predictions. In the case of critical Higgs inflation, however, there is no difference as far as we analyze the potential at field values close to the one of the inflection point, $\phi(t_i)$, as in this case $\Omega \approx 1$. From now on, we will follow prescription I for definiteness. Since we will be working around the inflection point, to avoid recursion problems in the numerical calculation we will take
the relation between $t$ (the argument of the running couplings) and $\phi(t)$ in eq. (\ref{eqU1c}) to be given by:
\beq
t=\ln \left(   \frac{ \alpha \bar \phi(t)}{ \sqrt{1+\xi(t_i) \bar \phi(t)^2 } }   \frac{M_P}{m_t} \right) \,\,.
\label{eq-tfinal}
\eeq
%with $\alpha=0.3$.

Before proceeding, it is important to understand the size of the theoretical error associated to the truncation of the effective potential at a certain loop order. This error can be estimated by varying $\alpha$, as done in the previous section. 
However, now that we apply this method to the model with the non-minimal coupling, we have also to take into account the effect of $\xi$. As far as $\xi$ is small, the plateau induced by it starts at higher field values than those of the inflection point, so that $U_i \approx V_i$. We can thus recover the results of fig.\,\ref{fig-Veff} (right panel), where we see that the better choice to reduce the logarithm is to take $\alpha$ in the range $0.1-1$: the tree-level potential displays a large variation with $\alpha$, but the 1-loop effective potential is reliable enough, in particular for the value $\alpha=0.3$, where it coincides with the 2-loop effective potential.

\begin{figure}[t!]
%\vskip .5cm 
\begin{center}
\includegraphics[width=16.5cm]{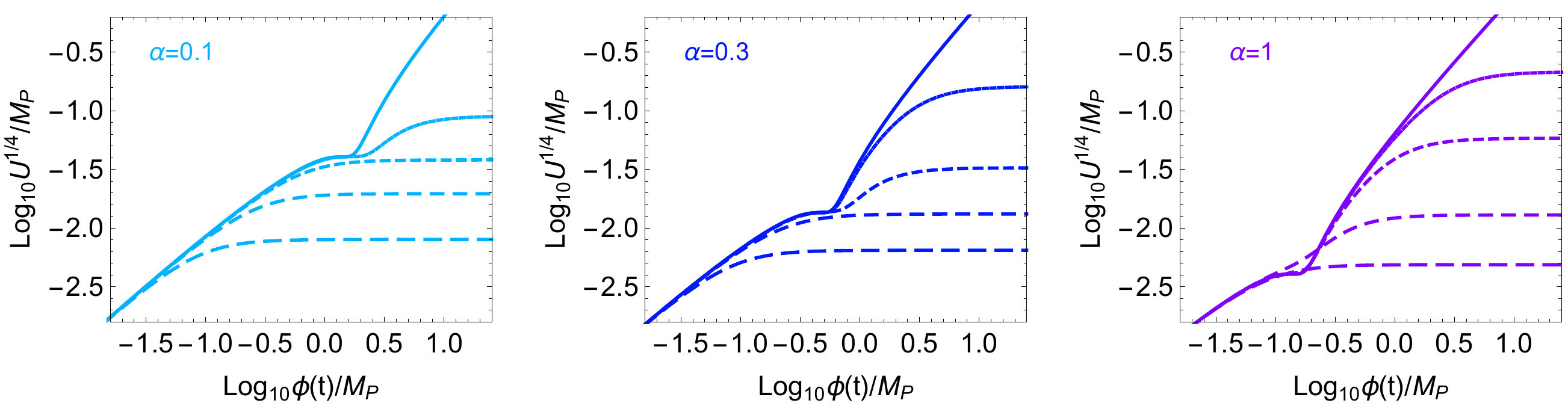}     
\vskip .4 cm
\includegraphics[width=16.3cm]{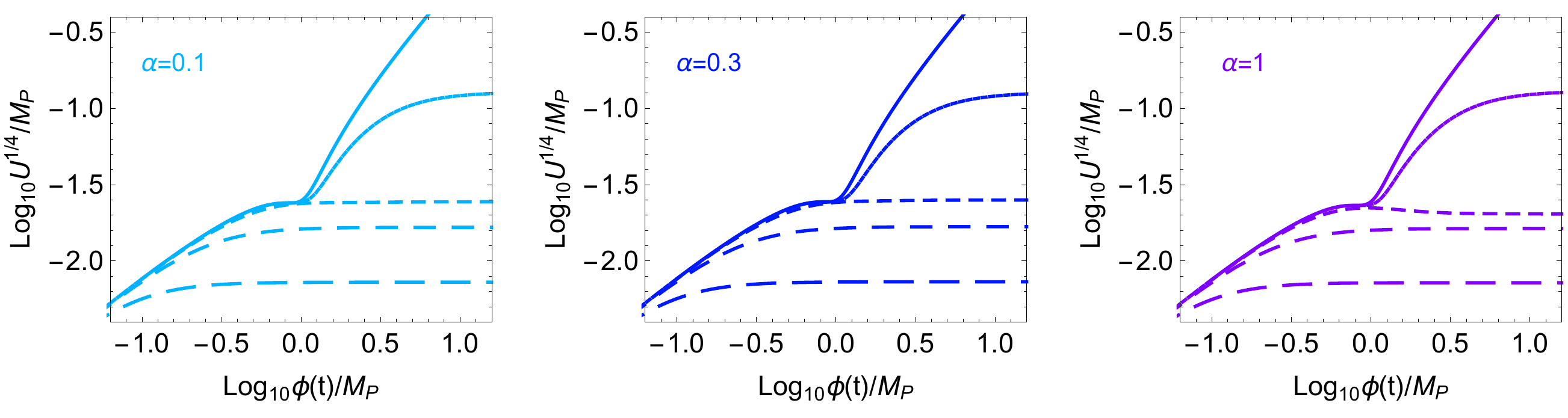}    
\end{center}
\caption{\baselineskip=15 pt  The RGE-improved potential $U$ is shown as a function of $ \phi(t)$ for $\alpha=0.1,0.3,1$, in the left, middle and right panel respectively. From top (solid) to bottom (long-dashed) the lines correspond to $\xi(0)=0,0.1,1,10,100$. Central values taken for $\alpha_s^{(5)}$ and $m_H$. Upper panel: tree-level calculation. Lower panel: 1-loop calculation. }
\label{fig-U-alfa}
\vskip .1 cm
\end{figure}

To see directly this, in the upper panel of fig.\,\ref{fig-U-alfa}, we display the tree-level potential $U^{(0)}$ as a function of $\bar \phi(t)$, taking $\alpha=0.1,0.3,1$, in the left, middle and right panel respectively. From top (solid) to bottom (long-dashed) the lines correspond to $\xi(0)=0,0.1,1,10, 100$. 
The central values are taken for $\alpha_s^{(5)}$ and $m_H$.
We can see that the highness of the inflection point is uncertain by one order of magnitude. The value of $\phi(t)$ where the inflection point occurs (a quantity that is not gauge invariant) is also quite undetermined: the same value of $\xi$ (e.g. $\xi(0)=10$) gives rise to a potential with an inflection point before the plateau for $\alpha=1$, while for $\alpha=0.1$ the plateau starts before the inflection point. This simply means that the tree-level potential, even improved with matching and running at NNLO, is not reliable. 

In the lower panel of fig.\,\ref{fig-U-alfa} we display the effective potential $U_{\rm eff}$ at 1-loop as a function of $\bar \phi(t)$, taking $\alpha=0.1,0.3,1$. We can see that these plots are essentially undistinguishable. This means that the 1-loop effective potential is trustable for the sake of the present analysis. 
In the following, we will thus consider the effective potential expansion at 1-loop and eq. (\ref{eq-tfinal}) taking 
$\alpha=0.3$ (for which the result of the 2-loop effective potential expansion is reproduced): in this way, the theoretical error associated to the truncation of the effective potential is smaller than the theoretical error associated to the matching of $\lambda$.

As a last step, we have to generalize the relation between $\phi$ and the canonical field $\chi$ to the case of running couplings. As discussed in refs. \cite{Espinosa:2015qea, Fumagalli:2016lls}, the kinetic term for the wave-function renormalized Higgs field, $\phi(t)$, can be made canonical by defining the field $\chi$ as:
%\begin{equation}
%\frac{d\chi}{d\phi(t)}= 
%\frac{\sqrt{1+\xi(t) (1+6\xi(t))  \bar \phi(t)^2}}{ 1+ \xi(t)  \bar \phi(t)^2}\,\,.
%\end{equation}
\begin{equation}
\frac{d \bar \chi}{d\bar \phi(t)}= \sqrt{K(t)}=
\frac{\sqrt{ 1+\xi(t) \bar \phi(t)^2 + 6 (\frac{1}{2} 
\frac{d\xi(t)}{d\bar\phi(t)} \bar \phi(t)+ \xi(t))^2   \bar \phi(t)^2} } { 1+ \xi(t)  \bar\phi(t)^2}
%\sqrt{\frac{\Omega^{2}+  \frac{3}{2} \left(   \frac{d\Omega^2}{d\phi} \right)^2 }{\Omega^{4}}}
\,,\qquad\bar \chi (\bar \phi(t)=0)=0\,\,.
\label{eq-dchidphit}
\end{equation}
We numerically integrate the equation above, substituting the argument $t$ of the running couplings as indicated in eq.\,(\ref{eq-tfinal}). 
%The closed analytical form of eq. (\ref{eq-closed}) can thus be generalized in a straightforward way to account for the running couplings:
%\beq
%\bar \chi( \bar \phi(t))= \sqrt{\frac{1+6\xi(t)}{\xi(t)}}\sinh^{-1}\left( \sqrt{\xi(t)(1+6\xi(t))} \bar \phi(t) \right) 
%-\sqrt{6} \tanh^{-1}\left(\frac{\sqrt{6}\xi(t) \bar \phi(t)}{\sqrt{ 1 +\xi(t)(1+6\xi(t)) \bar \phi(t)^{2}}}\right)    \,.
%\label{eq-closed}
%\eeq
In this way, we take into account the implicit dependence of $\xi(t)$ upon $\bar\phi(t)$ \cite{Espinosa:2015qea, Fumagalli:2016lls}. Note that we include also the term proportional to $d\xi(t)/ d\bar \phi(t)$ as suggested in ref.\,\cite{Ezquiaga:2017fvi}: the inclusion of such term is however numerically negligible. 
Actually,
we verified that also the approximation of taking $\xi(t)$ in eq. (\ref{eq-dchidphit}) constant and equal to the value it has at the inflection point, $\xi(t_i)$, is a very good approximation: the plots of the following section would not change.
%suggests another procedure to define the canonical field $\chi$ which,
%according to method proposed in \cite{Espinosa:2015qea, Fumagalli:2016lls} and followed here, is not justified in the context of the SM with non-minimal coupling to gravity. 

%Note that a large running of $\xi$ \cite{Bezrukov:2017dyv} was adopted in \cite{Ezquiaga:2017fvi}; we think that (despite the similar name), the model studied in ref. \cite{Ezquiaga:2017fvi} is de facto different from the one considered here and in \cite{Bezrukov:2017dyv}. 

\vskip 1.cm
%%%%%%%%%%%%%%%%%%%%%%%%
\section{The inflection point of Critical Higgs inflation }

We are now in the position to study in detail the potential corresponding to a critical configuration, first in terms of $\phi(t)$ and then expressing the potential as a function of the canonical field $\chi$, which is necessary to study the dynamics of inflation. 

The critical configuration is achieved when there is an inflection point at some field value $\phi_i \equiv \phi(t_i)$, and the plateau induced by the non-minimal coupling $\xi$ starts at a higher field value. %, $\phi(t_\xi) > \phi(t_i)$. 
The value of $\phi_i$ is fixed by the experimental window of the input parameters, namely $\alpha_s^{(5)}$ and $m_H$ ($m_t$ is chosen accordingly). The plateau instead starts when 
$ \bar \phi(t) \approx 1/\sqrt{\xi(t)}$; denoting by $t_\xi$ the renormalization scale where this happens, we define $\phi_\xi \equiv \phi(t_\xi)$.

The bottom central panel of fig.\,\ref{fig-U-alfa} shows that, taking $\alpha_s^{(5)}$ and $m_H$ at their central values, 
$\bar \phi_i \approx 1$ and $\bar U_i^{1/4} \approx 10^{-1.6}$ (namely $U_i^{1/4} \approx 6 \times 10^{16} $ GeV). 
The value of $\phi_i$ is not gauge invariant, but the highness of the potential at the infection point, $\bar U_i$, is (see the discussion in the previous chapter). 
Only with $\xi \lesssim 1$ one can have a critical configuration: with larger values the plateau destroys the inflection point.

Notice that, in a critical configuration, as far as we consider field values close to $\phi_i$, we have $\Omega \approx 1$. This has a two implications. 
Firstly, $U_i\approx V_i$ and we can apply here too all the discussion made in section 2.4.
Secondly, the relation between the renormalization parameter $t$ and $\phi(t)$ is the same for the two prescriptions, see eqs. (\ref{renorm1}) and (\ref{renorm2}): the value of $U_i$ is thus not plagued by the issue of the prescription (the behavior at the plateau actually is, but this will turn out to be not relevant for the sake of our discussion).

The presence of higher-dimensional operators close to the Planck scale might affect the critical configuration \cite{Bezrukov:2017dyv}. The small value of $\xi$ required for critical inflation is anyway particularly interesting, as it is related to one of the most significant drawbacks of Higgs inflation: the violation of perturbative unitarity at the scale $\bar \phi_U \approx 1/\xi$.  For $\xi < 1$ this scale is pushed at higher values
than the inflationary scale $\bar \phi_\xi$ and the assumptions of non-renormalizable
operators or new strong dynamics entering to restore unitarity are no longer required (see e.g. \cite{Allison}
and references therein).

\begin{figure}[t!]
%\vskip .5cm 
\begin{center}
\includegraphics[width=7.7 cm]{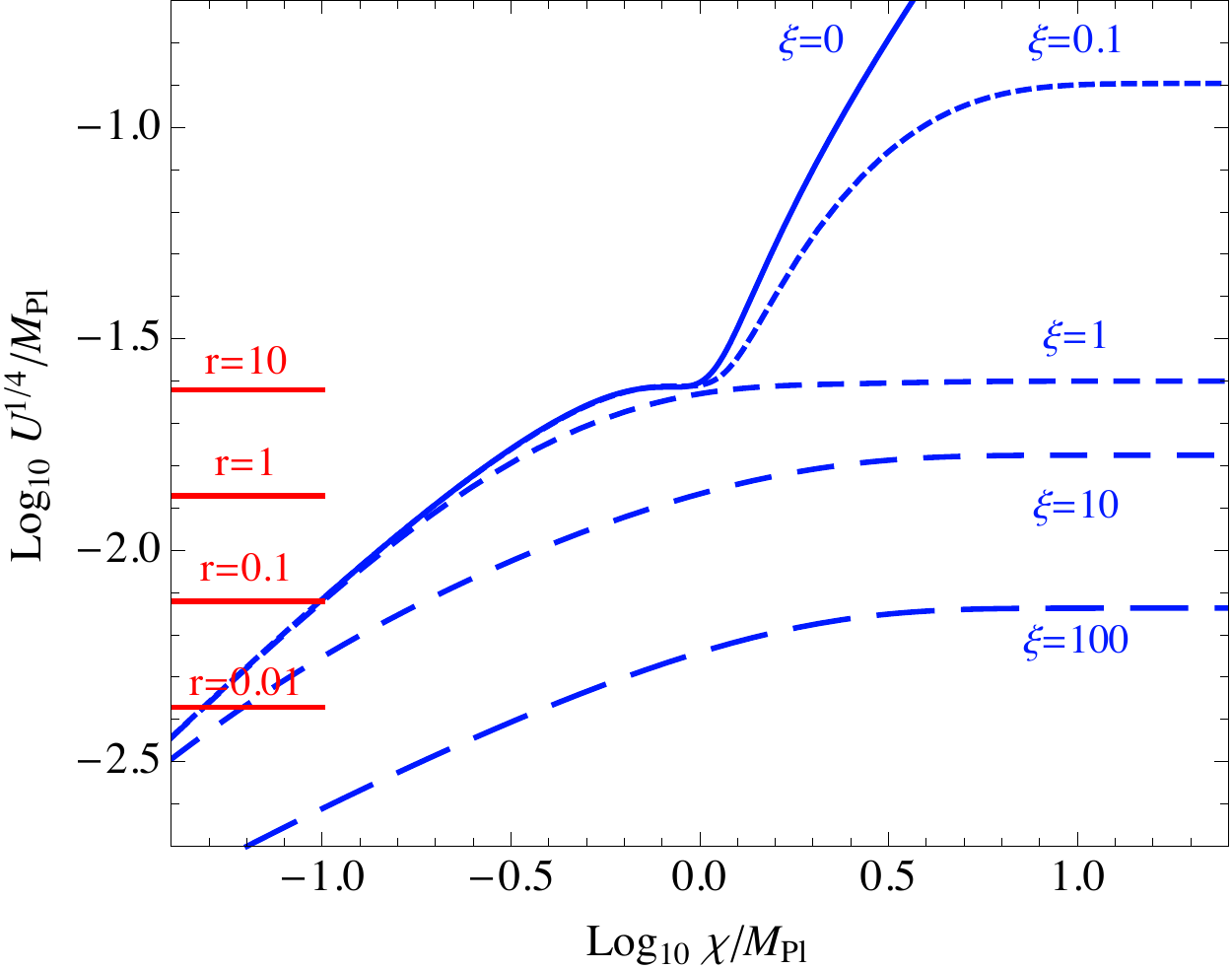}     \,\,\,\,\,
\includegraphics[width=7.7 cm]{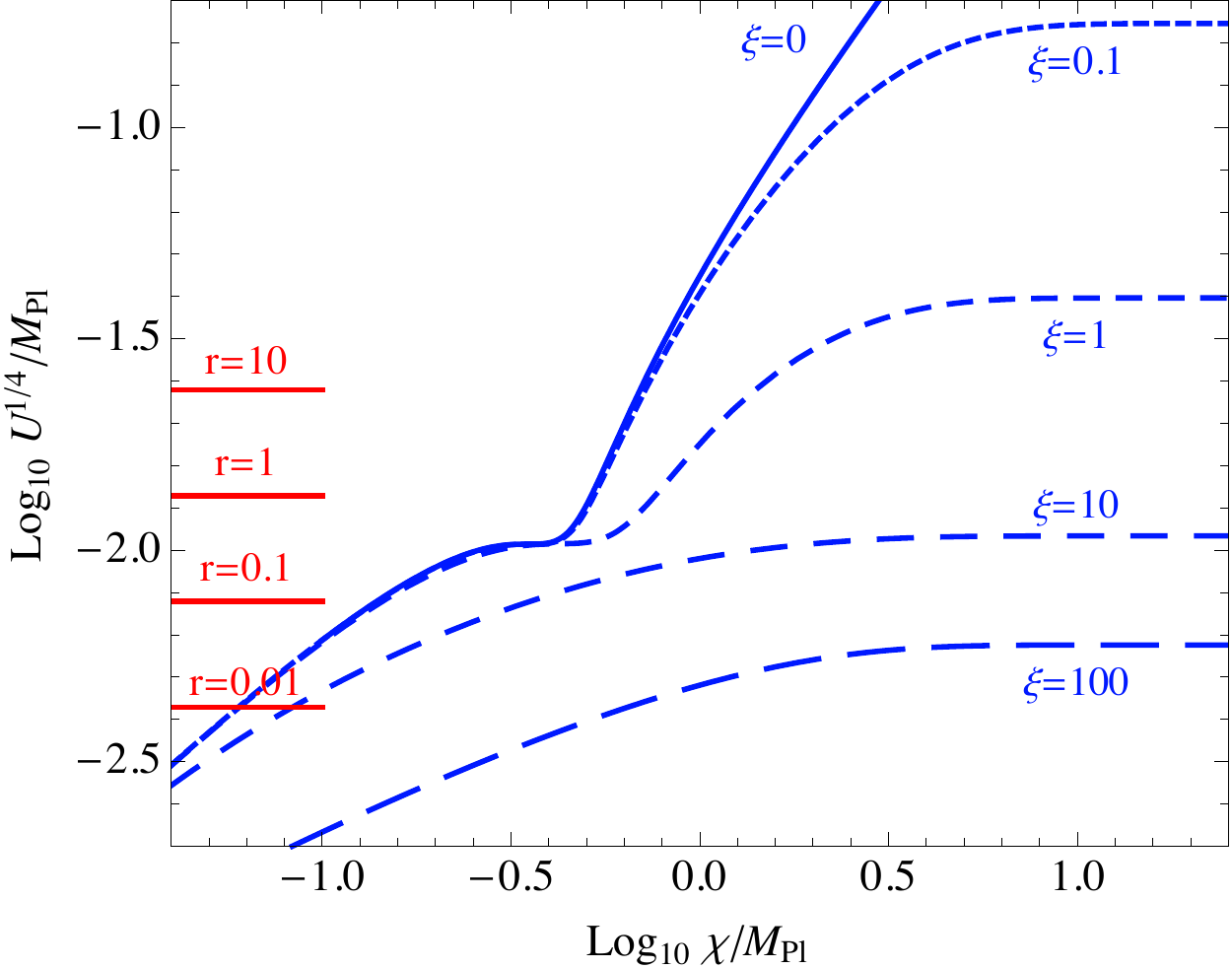}    
\end{center}
\caption{\baselineskip=15 pt The 1-loop effective potential $U$ is shown as a function of the canonical field $\chi$ (for $\alpha=0.3$). From top (solid) to bottom (long-dashed) the lines correspond to $\xi(0)=0,0.1,1,10, 100$. 
Left: Central values taken for $\alpha_s^{(5)}$ and $m_H$, $m_t=171.08$ GeV. 
Right: $\alpha_s^{(5)}$ is at its $3\,\sigma$ upper value, $m_H$ at its lower $1\,\sigma$, $m_t=172.08$ GeV.}
\label{fig-U1-sigma}
\vskip .1 cm
\end{figure}

We now turn to the field $\chi$, which allows to better inspect the dynamics of inflation. 
In the left panel of fig.\,\ref{fig-U1-sigma} we reproduce the same configuration shown in the bottom central panel 
of fig.\,\ref{fig-U-alfa}, obtained taking $\alpha_s^{(5)}$ and $m_H$ at their central values and $m_t=171.08$ GeV.
Clearly, the value of the effective potential at the inflection point does not change upon this substitution, as the relation between $\phi(t)$ and $\chi$ is a monotonically increasing one.
We see again that criticality, namely $\bar \chi_{\xi} > \bar \chi_i$, requires $\xi \lesssim 1$.

Once the shape of $U(\chi)$ is known, it is possible to calculate the inflationary observables.
Introducing the cosmological time $t$, the equation of motion of the field $\chi(t)$ is
\beq
\chi(t)'' +3 \,H(t)\, \chi(t) = - \frac{dU}{d\chi}(\chi(t)) \,\,, \,\, H(t)^2=  \frac{1}{3} \left(U(\chi(t)) + \frac{1}{2} \chi(t)^2 \right) \,\, ,
\eeq
where the initial conditions are $\chi(t_0)=\chi_0$, $\chi'(t_0)=\chi'_0$, and $t_0$ is some initial time.
The time duration of the inflationary phase is represented by the number of e-folds,
\beq
N=\int_{t_b}^{t_e} dt \, H(t)\,\,,
\eeq
where $t_e$ is the time of the end of inflation and $t_b>t_0$ is the time when the inflationary CMB observables, like $A_s, n_s, r$, are measured. It is known that $t_b$ is such that $N\approx 62$.

The critical configuration in the context of Higgs inflation \cite{Bezcritical, Hamada:2014iga, Hamada:2014wna}  has received interest in relation to the generation of primordial black holes \cite{Ezquiaga:2017fvi, Bezrukov:2017dyv}.
The general idea \cite{Ezquiaga:2017fvi, Garcia-Bellido:2017fdg, Kannike:2017bxn, Ballesteros:2017fsr} is that the inflaton field is slowly rolling on top of a plateau about $62$ e-folds before the end of inflation: CMB observables are measured at that epoch. %Initial conditions are not even relevant \cite{Salvio:2017oyf}.
About $20-30$ e-folds before the end of inflation, the inflaton crosses an inflection point where it slows considerably:  this would give rise to a peak in the power spectrum of primordial curvature perturbations, which would also result in a peculiar phenomenology for black holes, enhancing those that could significantly contribute to dark matter today.
Critical Higgs inflation \cite{Bezcritical, Hamada:2014iga, Hamada:2014wna} is indeed a nice and phenomenologically motivated realization of such scenario. We now study numerically how it could work.

In a critical configuration of Higgs inflation, the Hubble constant at the non-minimal plateau is higher than at the inflection point. Similarly to the discussion in the previous section, we can derive an upper bound on $U_i$ from the experimental upper bound on the tensor-to-scalar ratio $r$:
\beq
A_s= %P_s(N=62) \simeq 
\frac{H^2}{8 \pi^2 \epsilon} \left.  \right|_{N=62} 
\simeq  \frac{U}{24 \pi^2 \epsilon}\left.   \right|_{N=62} \simeq  \frac{2 \,U |_{N=62}}{3 \pi^2 r} \gtrsim \frac{2\,U_i}{3 \pi^2 r}\,\, ,
\eeq
where the last inequality holds because $U |_{N=62} \gtrsim U_{i} $. We thus have
\beq
% U_i \lesssim \frac{3 \pi^2}{2} r A_s \,\, .
 r \gtrsim  \frac{2}{3 \pi^2} \frac{U_i}{A_s} \,\, .
 \label{eq-bound}
\eeq
Since $U_i \approx V_i$ in a critical configuration, we can apply all the discussion made in section 2.4 for the inflection point of the SM.

So, without any further calculation, just looking at fig.\,\ref{fig-r}, we can conclude that, assuming the correct amplitude of scalar perturbations, for the present central values of  $\alpha_s^{(5)}$ and $m_H$, critical Higgs inflation would predict $r\gtrsim 10$; the dominant theoretical error in the calculation is the one associated to the matching of $\lambda$ and amounts to a factor of about $2$.
Even assuming that the theoretical error goes in the "right" direction of lowering $r$, in order to fulfill the present upper bound $r<0.12$ \cite{Ade:2015lrj, Ade:2015tva}, $\alpha_s^{(5)}$ should be set at its upper $3\,\sigma$ value and $m_H$ at its lower $1\,\sigma$ one.

We can see directly this tension looking at fig.\,\ref{fig-U1-sigma}, where the red horizontal segments show the values of $r$ 
according to the relation $r = 2 U_i/(3 \pi^2 A_s)$.
The plot in the left panel shows that, for the central values of $\alpha_s^{(5)}$ and $m_H$, the critical configuration predicts $r \gtrsim 10$. The present bound on $r$ implies that Higgs inflation is allowed only with $\xi \gtrsim  100$, hence far from criticality.

In the right panel of fig.\,\ref{fig-U1-sigma} we take $\alpha_s^{(5)}$ at its $3\,\sigma$ upper value and $m_H$ 
at its lower $1\,\sigma$ value: now we see that critical Higgs inflation would predict $r \gtrsim 0.3$. It would be possible to reduce the prediction down to $r \sim 0.12$ only invoking the theoretical error, and taking a suitable value for $\xi$. 
If one does not, the present bound on $r$ implies that Higgs inflation is allowed only with $\xi \gtrsim  10$, far from criticality.

Measuring $r$ close to its present upper bound would thus be compatible with Higgs inflation, but not in its critical version.
This would reasonably imply that the production of black holes during inflation is insufficient to constitute a significant fraction of the dark matter seen today.

%Our findings are thus different with respect to those of ref. \cite{Ezquiaga:2017fvi}.
For the sake of completeness, we now compare our findings with those of ref.\,\cite{Ezquiaga:2017fvi}: we think that, in addition to the large running of $\xi$ \cite{Bezrukov:2017dyv}, this work assumes a too small value for the quartic coupling $\lambda$ at the inflection point. The last analysis works at tree-level in the effective potential and finds that large black holes production and CMB observables require the value of $\lambda$ at the inflection point to be in the interval $(10^{-3} - 0.8) \times 10^{-6}$ (the Higgs potential being normalized as $V=1/4 \lambda \phi^4$). This range has to be compared with the one of our effective $\lambda$ at 1-loop: taking $\alpha_s^{(5)}$ to vary between its central and upper $3\,\sigma$ value, our effective $\lambda$ at 1-loop (normalized as in \cite{Ezquiaga:2017fvi}) rather spans the interval $(3.07-2.94)\times 10^{-6}$. 
%

%%%%%%%
\section{Discussion and conclusions}
\label{sec-concl}

We studied carefully the model of critical Higgs inflation \cite{Bezcritical, Hamada:2014iga, Hamada:2014wna}, calculating the Higgs effective potential according to the present state of the art, that is the NNLO. We found that, in order to satisfy the present upper bound on the tensor-to-scalar-ratio, $r<0.12$ \cite{Ade:2015lrj, Ade:2015tva}, while accounting for the correct amplitude of scalar perturbations, one should take $\alpha_s^{(5)}$ at its upper $4\,\sigma$ value, namely $\alpha_s^{(5)}=0.1233$. 
This tension can be alleviated at $3 \,\sigma$ by invoking the theoretical error (the dominant one is associated to the matching of $\lambda$) to go in the right direction.

Is $\alpha_s^{(5)}=0.1233$ too large? The current $1\,\sigma$ world average, $\alpha_s^{(5)}=0.1181 \pm 0.0013$ \cite{Patrignani:2016xqp}, is the result of a fit of many measurements:
those pointing to small values are the ones related to structure functions; lattice results also point towards small values, especially because the precision should be better than for other measurements; electroweak precision fits provide a larger error, so that 
$\alpha_s^{(5)}=0.1196 \pm 0.0030$. Anyway, looking at fig.\,9.2 of the PDG review on Quantum Chromodynamics \cite{Patrignani:2016xqp}, it seems quite unrealistic that $\alpha_s^{(5)}$ will turn out to be at the level of  $0.1233$. 

Assuming that the present $1\,\sigma$ world average of $\alpha_s^{(5)}$ will be confirmed in the future, 
one has to conclude that the model of critical Higgs inflation is in serious trouble {\it per se}, as it badly violates the present bound 
$r<0.12$ \cite{Ade:2015lrj, Ade:2015tva}. 
A fortiori, it is quite unrealistic that it might account for a significant fraction of the dark matter seen today under the form of primordial black holes. 

Unless $\alpha_s^{(5)}$ will turn out to be significantly larger than now estimated, two options are left:\\
1) Higgs inflation \cite{BezHiggs, Beztwoloop} is indeed the right model of primordial inflation, but it is realized in a non-critical form. Primordial black holes might be generated, but it is likely that they marginally contribute to the dark matter seen today;\\
2) the shape of the inflationary potential is indeed similar to the one of critical Higgs inflation, but $V_i$ is significantly lowered because of the effects of new physics. In principle, in this case primordial black holes might contribute to the dark matter \cite{Garcia-Bellido:2017fdg, Ballesteros:2017fsr, Kannike:2017bxn}. 
We checked that right-handed neutrinos would not help (as they have the same effect of enhancing the value of $m_t$). Maybe it would be more promising to introduce another scalar, but then the model would no more be of single field inflation, and the analysis would be accordingly more complicated.

%%%%%%%%%%%%%%%%%%%
\section*{\large Acknowledgements}

We thank the CERN Theory Department for kind hospitality and support during the completion of this work.
We acknowledge partial support by the research project TAsP (Theoretical Astroparticle Physics) funded by the Istituto Nazionale di Fisica Nucleare (INFN). We thank J. M. Ezquiaga, J. Fumagalli, J. Garcia-Bellido, E. Ruiz Morales, J. Rubio and M. Raidal for useful discussions.

%%%%%%%%%%%%%%%%%%%%%%%%%%%%%
\bibliographystyle{elsarticle-num} %stile bibliografico
\bibliography{bib} %qui richiamo il database
\end{document}